\documentclass[preprint,12pt]{elsarticle}
\usepackage{CJK}
\usepackage{graphicx}
\usepackage{amsmath}
\usepackage{epsfig}
\usepackage{indentfirst}
\usepackage[version=3]{mhchem}
\usepackage{float}

\usepackage{amssymb}

\begin{document}
\begin{CJK*}{GBK}{song}
\begin{frontmatter}



\title{Proton cloud and the possibility of direct perceiving of a Hydrogen nucleon}


\author{Li Yang}

\author{Ya-Qi Song}

\address{Institute of Information Engineering, Chinese Academy of Sciences, Beijing 10093, China}
\begin{abstract}
We introduce a concept of proton cloud and calculate the radius of the proton cloud of the Hydrogen atom. Then, we estimate the radius of the proton cloud of a Hydrogen atom on highly excited Rydberg states. Based on the size of proton cloud, the stability of the atom and technical level, we guess that the direct perceiving of the Hydrogen nucleon cloud, or proton cloud, is possible in near future.
\end{abstract}


\end{frontmatter}
\newtheorem{theorem}{Theorem}
\newtheorem{lemma}[theorem]{Lemma}
\newtheorem{conjecture}[theorem]{Conjecture}
\newtheorem{corollary}[theorem]{corollary}
\newtheorem{definition}{Definition}
\newtheorem{proposition}[theorem]{Proposition}
\newtheorem{remark}{Remark}
\newtheorem{protocol}{Protocol}



\section{Introduction}
\label{intro}
The nucleon of Hydrogen \ce{H_1^1} consists of one proton. The external dynamics of this proton is rather different from that of a free proton. The reason is that the proton belongs to a bound state with an electron.

The electron cloud is used to describe the state of electrons which circle around the nucleon of an atom. Born presented that to a state in space there corresponds a definite probability, which is given by the de Broglie wave associated with the state{\cite{Max1937}}. Schr\"{o}dinger Equation gives a dynamics to the de Broglie wave. Based on these, we could describe where electron is when it circles around the nucleon of a Hydrogen atom, which is called the electron cloud of a Hydrogen atom. The concept `electron cloud' makes our understanding of the atomic physics ocularly{\cite{Har1931}}. Here we investigate the concept of the nucleon cloud{\cite{Li2004}}. We believe it is interesting to perceive the nucleon cloud directly. The aim of the note is to demonstrate the probability of observing the nucleon cloud in nowadays laboratory.

\section{Proton Cloud}
\label{sec:1}
Because the Hydrogen nucleon belongs to a bound state of an electron and the nucleon, the external dynamical state of the nucleon is different from that of a free proton. Due to Coulomb reaction, the motion of nucleons in Hydrogen atoms is influenced by the motion of electrons. Therefore, the distribution and radius of proton cloud may have a relationship with those of electron cloud.

\subsection{Electron Cloud of Hydrogen atom}
\label{subsec:1}
For time-independent potentials, the Schr\"{o}dinger Equation of the bound states of the Hydrogen atom is:

\begin{eqnarray}\label{eq:1}
\begin{split}
&\bigg[-\frac{\hbar^2}{2m_1}(\frac{\partial^2}{\partial {x_1}^2}+\frac{\partial^2}{\partial {y_1}^2}+\frac{\partial^2}{\partial {z_1}^2})-\\
&\frac{\hbar^2}{2m_2}(\frac{\partial^2}{\partial {x_2}^2}+ \frac{\partial^2}{\partial {y_2}^2}+\frac{\partial^2}{\partial {z_2}^2})+U(x,y,z)\bigg]\Psi' =E_T\Psi',
\end{split}
\end{eqnarray}
where $(x_1, y_1, z_1)$ is the coordinate of the electron, $m_1$ is the mass of the electron, $(x_2, y_2, z_2)$ is the coordinate of the nucleon, $m_2$ is the mass of the nucleon, $E_T$ is the total energy of the system, and $U(x, y, z)$ is the potential energy of the system.

In two-particle system, it is convenient to solve the wave function in center-of-mass coordinate system. $(x, y, z)$ is the relative coordinate of the electron and the nucleon, then
\begin{eqnarray}\label{eq:2}
\begin{split}
&x=x_1-x_2,\\
&y=y_1-y_2,\\
&z=z_1-z_2.
\end{split}
\end{eqnarray}
$(X, Y, Z)$ is the coordinate of center-of-mass, and $M$ represents the total mass of the system. So we can get
\begin{eqnarray}\label{eq:3}
\begin{split}
&MX=m_1x_1+m_2x_2,\\
&MY=m_1y_1=m_2y_2,\\
&MZ=m_1z_1+m_2z_2.
\end{split}
\end{eqnarray}
Then, the Schr\"{o}dinger Equation becomes
\begin{eqnarray}\label{eq:4}
\begin{split}
&\bigg[-\frac{\hbar^2}{2M}(\frac{\partial^2}{\partial X^2}+\frac{\partial^2}{\partial Y^2}+\frac{\partial^2}{\partial Z^2})-\\&\frac{\hbar^2}{2\mu }(\frac{\partial^2}{\partial x^2}+\frac{\partial^2}{\partial y^2}+\frac{\partial^2}{\partial z^2})+U(x,y,z)\bigg]\Psi =E_T\Psi,
\end{split}
\end{eqnarray}
where
\begin{eqnarray}\label{eq:5}
\mu \equiv \frac{m_1m_2}{m_1+m_2},
\end{eqnarray}
is the reduced mass of the system.

By separation of variables, letting $\Psi(X,Y,Z;x,y,z)=\psi(x,y,z)\phi(X,Y,Z)$, Equation (\ref{eq:4}) could be separated into two independent equations:
\begin{eqnarray}\label{eq:6}
-\frac{\hbar^2}{2\mu }(\frac{\partial^2}{\partial x^2}+\frac{\partial^2}{\partial y^2}+\frac{\partial^2}{\partial z^2})\psi+U(x,y,z)\psi =E\psi,
\end{eqnarray}
\begin{eqnarray}\label{eq:7}
-\frac{\hbar^2}{2M}(\frac{\partial^2}{\partial X^2}+\frac{\partial^2}{\partial Y^2}+\frac{\partial^2}{\partial Z^2})\phi = (E_T-E)\phi.
\end{eqnarray}

Equation (\ref{eq:6}) is the description of wave function $\psi$ satisfying the relative motion between electron and proton, and energy $E$ of relative motion is the energy levels of the bound state of a proton and an electron. Equation (\ref{eq:7}) is the description of the wave function $\varphi$ satisfying the motion of the center-of-mass. And it is also the Schr\"{o}dinger Equation of a free particle with energy $E_T-E$, that is, the center-of-mass can be seen to move as free particle with energy $E_T-E$.

Through the wave function of the electron, where the electron is in space can be computed. Change Equation (\ref{eq:6}) into the expression of spherical coordinates. When the Hydrogen atom is in the state of $\Psi_{nlm}(r,\theta,\varphi)$, the probability of the electron appearing in the element volume of $d\tau = r^2\sin\theta drd\theta d\varphi$ around the point $(r,\theta,\varphi)$ is{\cite{shixun2009}}:
\begin{eqnarray}\label{eq:8}
W_{nlm}(r,\theta,\varphi)r^2\sin\theta drd\theta d\varphi = |\psi_{nlm}|^2r^2\sin\theta drd\theta d\varphi.
\end{eqnarray}
Then, the probability of the electron appearing in solid angle $d\Omega = \sin\theta drd\theta d\varphi$ in the direction of $(\theta,\varphi)$ can be written as:
\begin{eqnarray}\label{eq:9}
\begin{split}
W_{nlm}(\theta,\varphi)d\Omega &= \int_{r=0}^{\infty}|R_{nl}(r)Y_{lm}(\theta,\varphi)|^2r^2drd\Omega \\
&=|Y_m(\theta,\varphi)|^2d\Omega \\
&= N_{lm}^{2}[P_{l}^{m}(\cos\theta)]^2d\Omega,
\end{split}
\end{eqnarray}
where $R_{nl}$ is the radial function, $Y_{lm}$ is the spherical harmonics function, $P_{l}^{m}$ is the associated Legendre polynomials, and $N_{lm}$ satisfies the equation followed:
\begin{eqnarray}\label{eq:10}
N_{lm}=\sqrt{\frac{(l-|m|)!(2l+1)}{4\pi(l+|m|)!}}.
\end{eqnarray}
It is apparent that $W_{lm}$ is independent of $\varphi$, hence $W_{lm}$ has Z-axial-rotation symmetry.

\subsection{Proton Cloud of Hydrogen Atom}
\label{subsec:2}
Equation (\ref{eq:6}) depicts the wave function of Hydrogen atom. $(-x, -y, -z)$ is the relative coordinate of the nucleon and the electron. Substitute $(x, y, z)$ by $(-x, -y, -z)$ yields the wave function of proton:
\begin{eqnarray}\label{eq:11}
\begin{split}
&-\frac{\hbar^2}{2\mu }\bigg[\frac{\partial^2}{\partial (-x)^2}+\frac{\partial^2}{\partial (-y)^2}+\frac{\partial^2}{\partial (-z)^2}\bigg]\psi +U(-x,-y,-z)\psi \\
&= -\frac{\hbar^2}{2\mu }\bigg[\frac{\partial^2}{\partial x^2}+\frac{\partial^2}{\partial y^2}+\frac{\partial^2}{\partial z^2}\bigg]\psi +U(x,y,z)\psi = E\psi.
\end{split}
\end{eqnarray}

Equation (\ref{eq:11}) has the same form as Equation(\ref{eq:6}), thus the angular distribution and probability in space are identical with that of electrons. And the probability of the proton appearing in solid angle $d\Omega = \sin\theta drd\theta d\varphi$ in the direction of $(\theta,\varphi)$ is the same as that of electron. Therefore, the qualitative analysis of the three-dimensional images of both electron clouds and proton clouds at different quantum numbers are shown in Figure {\ref{fig:1}}:

\begin{figure}[H]
  \includegraphics[scale=0.4,bb= -150 0 550 600]{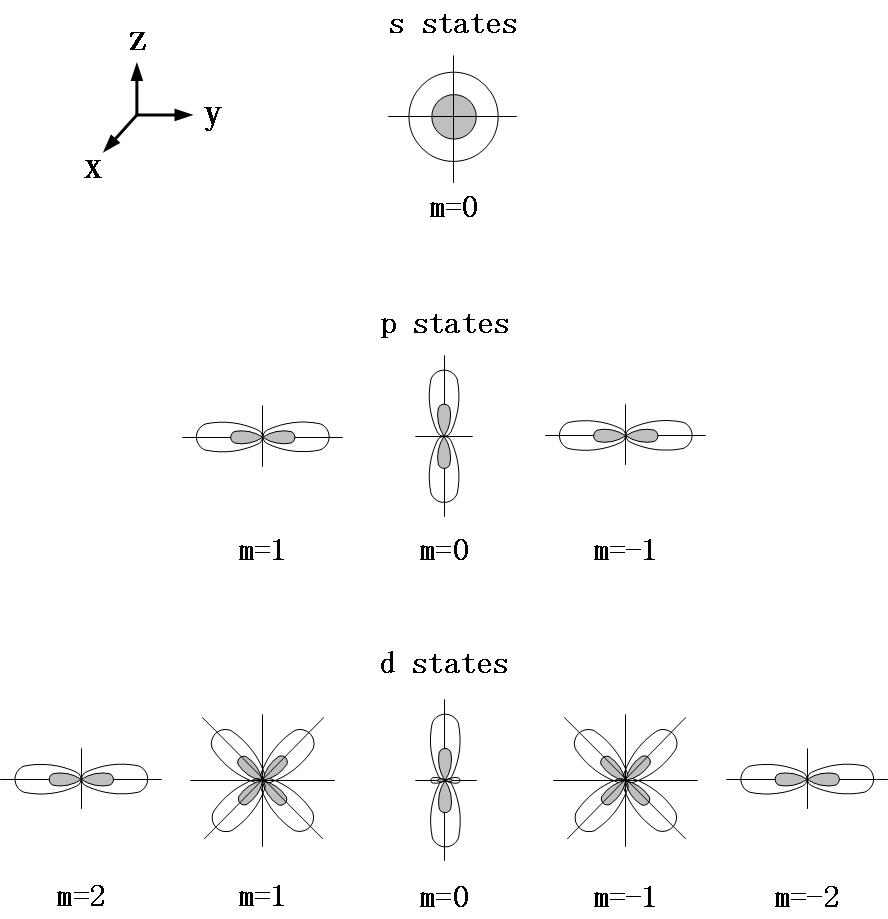}
\caption{The angular distribution of the electron cloud and the proton cloud of s,p,d quantum states.}
\label{fig:1}       
\end{figure}

From Equation (\ref{eq:2})(\ref{eq:3}) , we can get that:
\begin{eqnarray}\label{eq:12}
x_2=-\frac{m_1}{m}x,y_2=-\frac{m_1}{m}y,z_2=-\frac{m_1}{m}z,\Rightarrow r_2=\frac{m_1}{m}r,
\end{eqnarray}
where $r_2$ is the radius of the nucleon cloud. Therefore, the distance between the electron cloud and the proton cloud is rather long.

The angular distribution of the electron and proton is the same, thus the probability of the electron appearing in the range of the proton cloud is independent of the spherical harmonics function,
\begin{eqnarray}\label{eq:p}
P = \int\limits_0^{{r_2}} {R_{nl}^2(r){r^2}dr},
\end{eqnarray}
where $P$ is the probability of the electron appearing in the range of the proton cloud. We can get that in the different states of Figure {\ref{fig:1}}, $P$ is approximately equal to $2 \times {10^{-10}}$. Obviously, it is a small probability event that the electron obstructs the observation.

We can draw a conclusion that there is no overlap of the clouds for some states if we look them from a special direction. For example, we can take photos in the direction of z-axis when it is in the state of p and $m=\pm1$. However, in other directions, if we observe by rays of light, the wavelength of light is much longer than the proton cloud that will cause the diffraction. The method of electron scattering also can't be used to observe a single proton. The high states of the atoms are usually not stable, such as the Rydberg state. The techniques in current we know can not be used without damaging the unstable high states of the atoms when observing from other directions. In future, there may be some techniques can solve the problem. In conclusion, it is possible to observe the proton cloud in theory.

\section{Rydberg Hydrogen Atom}
\label{sec:3}
Rydberg atom is an atom in which an electron is excited to a high quantum state. Correspondingly, the orbit radius of electron in Rydberg state is fairly large, and that of proton cloud is even greater than the normal state. Then, Rydberg Hydrogen atoms offer much convenience to observe than ground state atoms. Energy levels $E_n$ of Hydrogen atoms and Hydrogen-like atoms/ions satisfy{\cite{Jean1998}}:
\begin{eqnarray}\label{eq:13}
E_n=E_{\infty}-\frac{R_mZ^2}{n^2},
\end{eqnarray}
where n stands for main quantum number, $E_{\infty}$ represents ionization energy, Z is nuclear charge number and $R_m$ is the Rydberg constant of mass m. Rydberg atoms have several properties:

(1)The volume of a Hydrogen atom in Rydberg states is large. The average orbit radius of the outer electron is approximately proportional to the square of quantum number n,
\begin{eqnarray}\label{eq:14}
\overline{r_{nl}}=n^2[1+\frac{1}{2}(1-\frac{l(l+1)}{n^2})]a_0,
\end{eqnarray}
where $l$ is the angular quantum number, while $a_0$ is the Bohr radius. When $n=630, l=0$, the radius is:
\begin{eqnarray}\label{eq:15}
\overline{r}=630^2\times0.53\times10^{-10}m\approx2.1\times10^{-5}m.
\end{eqnarray}
\\
From Equation (\ref{eq:12}), the radius of the proton cloud of the atom with $n=630$ is:
\begin{eqnarray}\label{eq:16}
r_2=\frac{r}{1836}\approx11.46nm.
\end{eqnarray}
Therefore, it is large enough for microscopy observation of the proton cloud.

(2)The lifetime of a Hydrogen atom in Rydberg states is long. According to quantum radiation theory, the spontaneous emission lifetime is approximately proportional to the cube of quantum number n, much longer than usual atom systems{\cite{Thomas2005}}, which facilitates the observation of the proton cloud.

The properties discussed above are the reasons that we suggest Hydrogen atom on highly excited Rydberg state as that in future experiment.

\section{Conclusion}
\label{sec:4}
In this note, the concept of the proton cloud is discussed. By investigating the relationship between the proton cloud and the electron cloud, we demonstrate that it is possible to observe the proton cloud from a specific direction. The radius of the proton cloud of a Hydrogen atom with $n=630$ is about 11.46nm which means the observation of the nucleon cloud in nowadays laboratory is possible. It can be seen that taking a photo of proton cloud is rather difficult, which involves the techniques of extremely cooling of Hydrogen atom and these have exact control of the state and orientation of the excited atom. However, these techniques have implemented by means of laser beams and external magnetizing fields. Based on the analysis given above, we guess that the direct perceiving of the Hydrogen atom nucleon cloud, or proton cloud, is possible in near future.

\bibliographystyle{model1a-num-names}
\bibliography{<your-bib-database>}









\end{CJK*}
\end{document}